\documentclass[twocolumn,aps,prl,showpacs]{revtex4}

\usepackage{graphicx}% Include figure files

\begin{document}
\title{Kochen-Specker Theorem for a Single Qubit Using Positive Operator-Valued Measures}
\author{Ad\'{a}n Cabello}
\email{adan@us.es}
\affiliation{Departamento de F\'{\i}sica Aplicada II,
Universidad de Sevilla, 41012 Sevilla, Spain}
\date{\today}
%First version: September 28, 2002.
%This version: May 6, 2003.
%After revising the second PRL's proofs.
%This will be quant-ph/0210082v2.

%%%%%%%%%%%%%%%%%%%%%%%%%%%%%% Abstract %%%%%%%%%%%%%%%%%%%%%%%%%%%%%%%

\begin{abstract}
A proof of the Kochen-Specker theorem for a single two-level
system is presented. It employs five eight-element positive
operator-valued measures and a simple algebraic reasoning based on
the geometry of the dodecahedron.
\end{abstract}

%%%%%%%%%%%%%%%%%%%%%%%%%%%%%%%%%%%%%%%%%%%%%%%%%%%%%%%%%%%%%%%%%%%%%%%%

\pacs{03.65.Ta,
%Foundations of quantum mechanics; measurement theory
03.65.Ud,
%Entanglement and quantum nonlocality
%(e.g. EPR paradox, Bell's inequalities, GHZ states, etc.)
03.67.-a}
%Quantum information
\maketitle

%%%%%%%%%%%%%%%%%%%%%%%%%%%%% Introduction %%%%%%%%%%%%%%%%%%%%%%%%%%%%%

It is a widely held belief that ``a single qubit is not a truly
quantum system in the sense that its dynamics and its response to
measurements can all be mocked up by a classical hidden-variable
model. There are no Bell inequalities or a Kochen-Specker theorem
for a two-dimensional system that forbids the existence of a
classical model''~\cite{vanEnk00}. Any proof of Bell's theorem of
the impossibility of local hidden-variables in quantum mechanics
requires a composite system. On the other hand, the standard proof
of the Kochen-Specker (KS) theorem~\cite{Specker60,Bell66,KS67} of
the impossibility of noncontextual hidden-variables applies only
to physical systems described by Hilbert spaces of dimension three
or higher.

In this Letter, I show that it is possible to extend the KS
theorem to a single two-level system (qubit). The key is to
consider generalized measurements, represented by positive
operator-valued measures (POVMs)
\cite{Neumark43,Ludwig76,Davies76,Peres93,NC00}, instead of just
standard measurements, represented by von Neumann's
projection-valued measures. This shall lead to a generalization of
the KS theorem which rules out all hidden-variable theories ruled
out by the original KS theorem plus some hidden-variable theories
for a single qubit. The common feature of all these
hidden-variable theories is that they are {\em noncontextual},
that is, they assign predefined yes-no answers to a set of
questions~$\{Q,R,S\ldots\}$ (tests) independently of whether
question~$Q$ is formulated jointly with question~$R$ or with a
different question~$S$. The physical content of the KS theorem can
also be summarized by saying that ``measurements'' do not reveal
preexisting values, or that ``unperformed experiments have no
results''~\cite{Peres78}.

The standard proof of the KS theorem is based on the observation
that, for a physical system described by a Hilbert space of
dimension~$d \ge 3$, it is possible to find a set of~$n$
projection operators, which represent yes-no questions about the
physical system, so that none of the~$2^n$ possible sets of
``yes'' or ``no'' answers is compatible with the sum rule for
orthogonal resolutions of the identity (i.e., if the sum of a
subset of mutually orthogonal projection operators is the
identity, one and only one of the corresponding answers ought to
be yes)~\cite{Peres93,Bub97}. The original proof required~$117$
projection operators in~$d=3$~\cite{KS67}. The actual record
stands at~$18$ projection operators in~$d=4$~\cite{CEG96}. The
record for~$d=3$ stands at~$31$ projection operators~\cite{CK90}.

It is impossible to prove the KS theorem for a single qubit
(described by a Hilbert space of~$d=2$) by using a set of
projection operators. Any proof of this kind would require that
any projection operator of the set be orthogonal to at least two
other projection operators. However, in~$d=2$ any projection
operator is orthogonal only to one projection operator. Therefore,
for~$d=2$, it is possible to assign yes and no answers to all
projection operators, satisfying the sum rule for orthogonal
resolutions of the identity. This explains why it is possible to
construct explicit noncontextual hidden-variable models that are
capable of reproducing all the predictions of quantum mechanics
{\em for von Neumman's measurements} on a single qubit
\cite{Bell66,KS67,BB66,Clauser71,Selleri90}.

Motivated by the quantum information approach to quantum mechanics
and by the fact that current technology allows an exquisite level
of control over the measurements that can be performed, recent
formulations of the principles of quantum mechanics
\cite{Peres93,NC00,Fuchs02} stress that the measurements
correspond to POVMs, extending the notion of von Neumann's
projection-valued measures. The main difference between POVMs and
von Neumann's projection-valued measures is that for POVMs the
number of available outcomes of a measurement may be higher than
the dimensionality of the Hilbert space. An~$N$-outcome
generalized measurement is represented by an~$N$-element POVM
which consists of~$N$ positive-semidefinite operators~$\{E_d\}$
that sum the identity (i.e.,~$\sum_d E_d= 1\!\!\:\!{\rm{I}}$).
Neumark's theorem~\cite{Neumark43} guarantees that there always
exists a realizable experimental procedure to generate any desired
POVM. Any generalized measurement represented by a POVM can be
seen as a von Neumann's measurement on a larger Hilbert space.
Therefore, any generalized measurement on a single qubit can be
seen as a von Neumann's joint measurement on a system composed by
the qubit plus an auxiliary quantum system (ancilla)~\cite{comm2}.
If we define the ancilla as belonging to the measuring apparatus,
then we can legitimately speak of a (generalized) measurement on a
single qubit~\cite{Peres93b}.

It has been shown that, when one considers an ancilla, then
noncontextual hidden-variables cannot reproduce the predictions of
quantum mechanics for von Neumman's measurements on pre- and
post-selected systems~\cite{Cabello97}. On the other hand, the KS
theorem can be seen as a consequence of Gleason's theorem
\cite{Gleason57}. Recently, a Gleason-like theorem using POVMs has
been proved~\cite{Busch99,CFMR00}. Unlike Gleason's theorem, the
new one is also valid for~$d=2$. This suggests that the KS theorem
could be extended to~$d=2$ by using POVMs instead of von Neumann's
measurements~\cite{Fuchs02}.

Physically, this would mean that it is impossible to construct a
noncontextual hidden-variables theory for a single qubit which
assigns an outcome, for instance~${\cal E}_A$, regardless of
whether this outcome belongs to a the POVM represented by~${\cal
E}_A$,~${\cal E}_B$, \ldots,~${\cal E}_M$ or to the POVM
represented by~${\cal E}_A$,~${\cal E}_b$, \ldots,~${\cal E}_m$.
This bears a close similarity to the original formulation of the
KS theorem~\cite{KS67} based on von Neumann's measurements on a
single spin-1 system. KS considered measurements of the type
\begin{equation}
H(x,y,z) = a S_x^2 + b S_y^2 + c S_z^2,
\end{equation}
where~$a$,~$b$, and~$c$ are real distinct numbers and~$S_x^2$ is
the square of the spin component along the~$x$ direction. A
measurement of~$H$ has three possible outcomes: $b+c$,~$a+c$, and
$a+b$. KS showed that whichever outcome actually occurs was not
predefined. They accomplished this by considering alternative
measurements~$H(x,j,k)$, where~$x$,~$j$, and~$k$ are mutually
orthogonal directions, and assuming that if the outcome of
measuring~$H(x,y,z)$ had/had not been~$b+c$, then the outcome of
measuring~$H(x,j,k)$ would have/have not been~$b+c$.

The challenge is to prove the KS theorem for~$d=2$ using
generalized measurements, that is, to find an explicit set of
POVMs on a single qubit so that none of the possible sets of~$2^n$
yes or no answers (where~$n$ is the number of different
positive-semidefinite operators in the POVMs) is compatible with
the sum rule for positive-semidefinite operators of a POVM (i.e.,
if the sum of a subset of positive-semidefinite operators is the
identity, one and only one of the corresponding answers ought to
be yes). Fuchs has suggested using sets of three-outcome POVMs of
the ``Mercedes-Benz'' type~\cite{Fuchs02}. So far, however, no
proof with this or any other type of POVMs has been described.

Let us define the following eight-outcome generalized measurement
on a qubit represented by the following eight-element POVM:~$\{
{\cal E}_{C+}, {\cal E}_{C-}, {\cal E}_{E+}, {\cal E}_{E-}, {\cal
E}_{F+}, {\cal E}_{F-}, {\cal E}_{G+}, {\cal E}_{G-} \}$, where
\begin{eqnarray}
{\cal E}_{C+} & = & {1 \over 4} P_{\neg |C=-1\rangle} =
{1 \over 4} \left(1\!\!\:\!{\rm{I}}-P_{|C=-1\rangle}\right)
\nonumber \\
& = & {1 \over 4} |C=+1\rangle\langle C=+1|,
\label{POVM1} \\
{\cal E}_{C-} & = & {1 \over 4} P_{\neg |C=+1\rangle} =
{1 \over 4} \left(1\!\!\:\!{\rm{I}}-P_{|C=+1\rangle}\right)
\nonumber \\
& = & {1 \over 4} |C=-1\rangle\langle C=-1|,
\label{POVM2}
\end{eqnarray}
and analogously~${\cal E}_{E+}$, etc.~$C$,~$E$,~$F$, and~$G$ are
the directions obtained by connecting the center of a cube with
its four nonantipodal vertices.~$P_{\neg |C=-1\rangle}$ is the
projection on the qubit states orthogonal to~$|C=-1\rangle$
(which, for example, could be the spin state along direction~$C$
with eigenvalue~$-1$ of a spin-1/2 particle). As can be easily
checked, the sum of these eight positive-semidefinite operators is
the identity.

%%%%%%%%%%%%%%%%%%%%%%%%%%%% Figure 1 %%%%%%%%%%%%%%%%%%%%%%%%%%%%%

\begin{figure}
\centerline{\includegraphics[width=7.6cm]{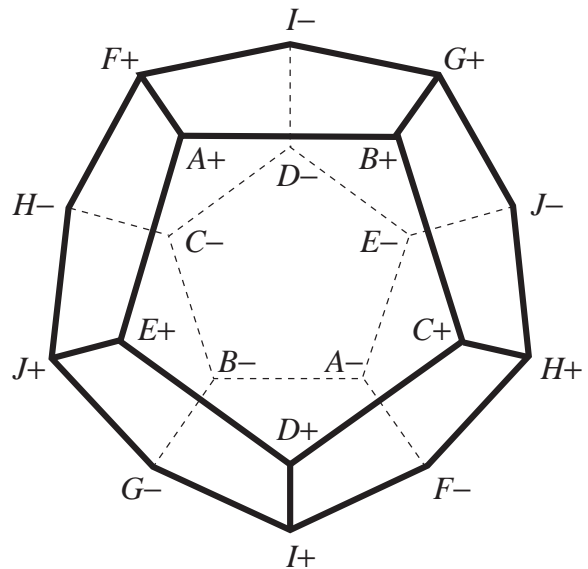}}
\caption{\label{dode01} Notation for the~$20$ vertices of the
dodecahedron:~$A+$ is the antipode of~$A-$.}
\end{figure}

%%%%%%%%%%%%%%%%%%%%%%%%%%%% Figure 2 %%%%%%%%%%%%%%%%%%%%%%%%%%%%%

\begin{figure}
\centerline{\includegraphics[width=6.5cm]{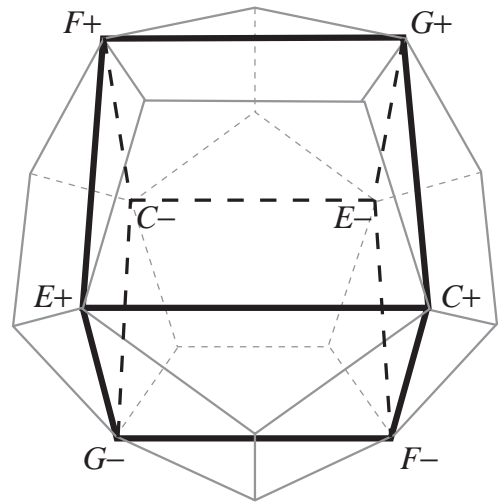}}
\caption{\label{dode02} The cube~$CEFG$ is one of the five
inscribed (sharing vertices) in the dodecahedron. It corresponds
to an eight-element POVM.}
\end{figure}

%%%%%%%%%%%%%%%%%%%%%%%%%%%%%%%%%%%%%%%%%%%%%%%%%%%%%%%%%%%%%%%%%%%%

Now let us consider the ten directions obtained by connecting the
center of a dodecahedron with its ten nonantipodal vertices,
labelled~$A$,~$B$, \ldots,~$J$ as in Fig.~\ref{dode01}. There are
only five cubes inscribed (sharing vertices) in a dodecahedron
(Fig.~\ref{dode02}). All of them share the same center, and any
two cubes share two antipodal vertices. Each cube allows us to
define an eight-element POVM similar to the one defined above. The
resulting five POVMs can be expressed as:
\begin{eqnarray}
{\cal E}_{A+}+{\cal E}_{A-}+{\cal E}_{C+}+{\cal E}_{C-}+ & & \nonumber
\\
{\cal E}_{I+}+{\cal E}_{I-}+{\cal E}_{J+}+{\cal E}_{J-} &=& 1\!\!\:\!{\rm{I}},
\label{GM1}
\\
{\cal E}_{A+}+{\cal E}_{A-}+{\cal E}_{D+}+{\cal E}_{D-}+ & & \nonumber
\\
{\cal E}_{G+}+{\cal E}_{G-}+{\cal E}_{H+}+{\cal E}_{H-} &=& 1\!\!\:\!{\rm{I}},
\label{GM2}
\\
{\cal E}_{B+}+{\cal E}_{B-}+{\cal E}_{D+}+{\cal E}_{D-}+ & & \nonumber
\\
{\cal E}_{F+}+{\cal E}_{F-}+{\cal E}_{J+}+{\cal E}_{J-} &=& 1\!\!\:\!{\rm{I}},
\label{GM3}
\\
{\cal E}_{B+}+{\cal E}_{B-}+{\cal E}_{E+}+{\cal E}_{E-}+ & & \nonumber
\\
{\cal E}_{H+}+{\cal E}_{H-}+{\cal E}_{I+}+{\cal E}_{I-} &=& 1\!\!\:\!{\rm{I}},
\label{GM4}
\\
{\cal E}_{C+}+{\cal E}_{C-}+{\cal E}_{E+}+{\cal E}_{E-}+ & & \nonumber
\\
{\cal E}_{F+}+{\cal E}_{F-}+{\cal E}_{G+}+{\cal E}_{G-} &=& 1\!\!\:\!{\rm{I}}.
\label{GM5}
\end{eqnarray}
Each equation contains eight positive-semidefinite operators whose
sum is the identity. Therefore, a noncontextual hidden-variable
theory must assign the answer yes to one and only one of these
eight operators. However, such an assignment is impossible, since
each operator appears twice in (\ref{GM1})--(\ref{GM5}), so that
the total number of yes answers must be an even number, while the
number of POVMs, and thus the number of possible yes answers, is
five.

Geometrically, this proof expresses the impossibility of coloring
black (for yes) or white (for no) the vertices of a dodecahedron
in such a way that each of the five inscribed cubes has one and
only one vertex colored black.

To my knowledge, this is the first proof of the KS theorem for a
single qubit. Moreover, this proof joins the three most wanted
features in a proof of the KS theorem: (a) It is based on a simple
algebraic argument (parity, like the proofs in
\cite{CEG96,Kernaghan94}), so checking the impossibility of
coloring requires neither an intricate geometrical argument
\cite{KS67} nor a computer program~\cite{Peres93}; (b) it needs
few operators (five POVMs containing only~$20$ different
positive-semidefinite operators); (c) it admits an elegant
geometrical interpretation. Curiously indeed, this interpretation
is in terms of the dodecahedron, whose geometrical properties were
also used in Penrose's proofs of the KS theorem
\cite{Penrose00,ZP93,Penrose94}.

%%%%%%%%%%%%%%%%%%%%%% Acknowledgements %%%%%%%%%%%%%%%%%%%%%%

I never met Rob Clifton personally, but for years we maintained
some correspondence on the KS theorem. He once told me:
``Kochen-Specker problems can be addictive and I am an addict!''
\cite{Clifton96}. I share this addiction. I would like to dedicate
this Letter to his memory. I would like to thank J. Bub, C. A.
Fuchs, and A. Peres for comments, and the Spanish Ministerio de
Ciencia y Tecnolog\'{\i}a Grants No.\ BFM2001-3943 and No.\
BFM2002-02815, and the Junta de Andaluc\'{\i}a Grant No.\ FQM-239
for support.

{\em Note added.---}After reading an earlier version of this
Letter, Masahiro Nakamura has found a simpler proof of the KS
theorem for a single qubit: Let~$A$,~$B$, and~$C$ be the three
directions obtained by joining the center of a regular hexagon
with its three nonantipodal vertices. A simple parity argument
shows that it is impossible to assign noncontextual yes-no answers
to the six positive semidefinite operators contained in the three
four-element POVMs~$\{ {\cal E}_{A+}, {\cal E}_{A-}, {\cal
E}_{B+}, {\cal E}_{B-} \}$,~$\{ {\cal E}_{B+}, {\cal E}_{B-},
{\cal E}_{C+}, {\cal E}_{C-} \}$, and~$\{ {\cal E}_{A+}, {\cal
E}_{A-}, {\cal E}_{C+}, {\cal E}_{C-} \}$, where \mbox{${\cal
E}_{A+}={1 \over 2} |A=+1\rangle\langle A=+1|$,}
etc.~\cite{Nakamura03}.

%%%%%%%%%%%%%%%%%%%%%%%%% References %%%%%%%%%%%%%%%%%%%%%%%%%


\begin{thebibliography}{99}
\bibitem{comm2}
Since a generalized measurement represented by POVMs can only be
implemented by means of a von Neumann's measurement on a higher
dimensional system, and since the standard KS theorem shows that
there cannot be a noncontextual hidden-variables theory for the
higher dimensional system, then one might think that the
impossibility of a noncontextual hidden-variables theory for the
lower dimensional system follows from the impossibility of a
noncontextual hidden-variables theory for the higher dimensional
system. However, there is a need for a proof using POVMs similar
to the one introduced in this Letter, because at present it has
not been proved that a standard proof of the KS theorem in the
higher dimensional system would lead to a proof of the KS theorem
using POVMs in the lower dimensional system.
\bibitem{Peres93b}
According to Peres, ``measurements are processes in which an
apparatus interacts with the physical system under study, in such
a way that a property of that system affects a corresponding
property of the apparatus''~\cite{Peres93}. The approach to the KS
theorem using POVMs is fully consistent with this definition: here
the physical system under study is a single qubit and the
apparatus includes the ancilla.
\end{thebibliography}
\end{document}